\documentclass[prb,showpacs,showkeys,preprintnumbers,superscriptaddress,a4paper,twocolumn]{revtex4-1}
\usepackage{graphicx}
\usepackage{subcaption}
\usepackage{float}
\usepackage{esvect}
\usepackage{bbold}
\usepackage[T1]{fontenc}
\usepackage[utf8]{inputenc}
\usepackage{graphicx}
\usepackage{amsmath,amssymb}
\usepackage{amstext}	
\usepackage{oldgerm}	
\usepackage[colorlinks,citecolor=blue,filecolor=blue,linkcolor=blue,urlcolor=blue]{hyperref}

\def\beq{\begin{equation}}
\def\eeq{\end{equation}}
\def\bea{\begin{eqnarray}}
\def\eea{\end{eqnarray}}
\def\ba{\begin{array}}
\def\ea{\end{array}}

\begin{document}
\title {Interacting fermions in two dimension in simultaneous presence
 of disorder and magnetic field}
\author{Saptarshi Mandal}\email{saptarshi@iopb.res.in} 
\affiliation{ Institute of Physics, Bhubaneswar 751005, Odisha, India}
\affiliation{Homi Bhabha National Institute, Training School Complex, Anushaktinagar, Mumbai 400094,  India}
\author{Sanjay Gupta}\email{guptajay2@gmail.com}
\affiliation{ Department of Physics, Central University of Jharkhand, India}
                   
\begin{abstract}

We study the  revival of Hofstadter butterfly due to the competition
between disorder and
electronic interaction using  mean field approximation  of unrestricted Hartree 
Fock method at zero temperature for two dimensional square and honeycomb lattices. 
Interplay of disorder and electronic correlation to nullify each other
is corroborated by the fact that honeycomb lattice needs more strength
of electronic correlation owing to its less co-ordination number which
enhances the effect of disorder.  The extent of revival of the butterfly is 
better in square lattice than honeycomb lattice due to higher coordination number. The effect of disorder and
interaction is also investigated to study entanglement entropy and
entanglement spectrum. We find that  for honeycomb lattice area law of entanglement entropy is obeyed in all cases but for square
lattice there is some departure from area law for larger subsystems.  The entanglement spectrum have the reflection symmetry
of the original butterfly of the Hofstadter spectrum. The interaction induces a gap in the
entanglement spectrum as well conforming the correspondence between physical spectrum and entanglement spectrum.
The effect of disorder closes the interaction induced gap in the entanglement spectrum establishing
the nullification  of interaction due to disorder and vice versa.
\end{abstract}

\date{\today}

\maketitle

\section{Introduction} Since its discovery in 1976 Hofstadter butterfly $\cite{Hofs}$ has remained a topic 
of great interest in condensed matter physics  and is still drawing attention for  studying  its various 
aspects. Hofstadter studied the properties of an electron moving in a two dimensional square lattice in the presence 
of an uniform perpendicular magnetic field. The spectrum when plotted as a function of the ratio $\phi/\phi_{0}$ 
of magnetic flux per plaquette to the flux quantum resembled a butterfly like structure and hence the name. 
The energy spectrum is an intricate function of this ratio $\phi/\phi_{0}$. If $\phi/\phi_{0} = p/q$ is a rational 
number then each energy band is devided into $q$ sub-bands by the magnetic field. When the magnetic field is varied, 
the spectrum shows a recursive structure at rational values of $p/q$, while  a Cantor set like structure appears at irrational
values of $p/q$ . The gaps appear because of Landau levels and  effect of interaction has not been considered.
Later on, the Hofstadter butterfly has been studied for different lattice structures like triangular $\cite{Yeong}$, 
honeycomb $\cite{Gumbs}$ and comparative studies of triangular and kagome lattices $\cite{Juan}$ etc. Experimental 
realization of Hofstadter butterfly has been achieved in  recent past $\cite{Dean}$. The work of Wilkinson $\cite{Wilkinson}$ based on semi classical approach  followed by re-normalization group scheme applied to energy spectrum provided theoretical insight into the Hofstadter butterfly.\\

\indent In practice we have both disorder and interaction simultaneously present in a physical system. In addition, study of disorder and interaction  are of genuine  theoretical interest for the complexity they bring in to an otherwise non-interacting translationally invariant system. Studies of effect of disorder  $\cite{Bhatt,koshino}$ and interaction $\cite{Minh}$  on Hofstadter butterfly have been carried out separately. Presence of disorder 
into such a system changes the scenario dramatically. Large disorder (large compared to hopping integral) completely destroys
the butterfly structure $\cite{Bhatt}$. It smears out  the butterfly structure by plugging in states into the gaps. 
For small disorder(small or comparable to hopping integral) it is expected  that states will appear in the gaps between 
sub bands thus smearing out the sub bands while the large gaps will not be affected $\cite{Bhatt}$. On the other hand effect 
of electronic correlation on the Hofstadter butterfly has been studied using exact diagonalization $\cite{Cjazka}$, self consistent 
Hartree approximation $\cite{Gudmundsson}$ and dynamical mean field theory(DMFT) $\cite{Minh}$. The exact diagonalization 
calculation for finite size system $\cite{Cjazka}$ showed that electronic correlation smear out the fine structure of Hofstadter 
butterfly. The self consistent Hartree approximation done earlier $\cite{Gudmundsson}$ showed that  in presence of 
electron correlation the pattern of spectrum heavily depends on the filling or chemical potential of the system. The effect of interaction on Hofstadter
butterfly has been studied using DMFT and other technique in case of Falikov-Kimball model $\cite{Minh,wrobel-2009,subhashree}$. It showed that in long range ordered phase, the presence of electronic interaction induced a gap without disturbing the fine structure of the butterfly. The gap simply
eats up the central part of the butterfly keeping intact the two wings. In the present work we deal with the issue of revival of 
the Hofstadter butterfly in simultaneous presence of site disorder and correlation. We incorporate the electronic correlation 
by considering the single orbital Hubbard model with onsite Coulomb repulsion. The site disorder has been introduced by wrapping 
the four letter Rudin-Shapiro sequence over all the sites of  lattice. The reason for choosing Rudin-Shapiro over other 
quasi-periodic sequence is that the dc conductivity calculated with it as site disorder for a non-interacting electron problem in 
one dimension was found to be in much better agreement with that obtained with random disorder compared to Fibonacci and Thue-Morse sequences$\cite{Aldea}$. 
As such one is able to arrive at results close to that of random disorder using a deterministic 
sequence and are saved from doing configurational average. The effect of disorder is to localize electrons. As a result sites with 
lowest onsite potential  have highest occupation of electrons and vice versa. On the other hand the on-site Hubbard interaction drives the
system towards an anti-ferromagnetic configuration which is also an insulating state $\cite{Gupta}$. Recently it is shown that interaction also induces charged ordering in the Hofstadter butterfly ~\cite{archana-2016,archana-2017}, however for Bose-Hubbard model in the presence of magnetic field such charge density order is absent ~\cite{umucahlar,oktel}.  On varying the on-site Hubbard interaction
in presence of disorder, the system therefore undergoes a transition from disorder driven insulator to correlation driven insulator. 
In between, at a certain critical value of the on-site Coulomb repulsion, the two  effects nullify each other. As such we expect a 
revival of the Hofstadter butterfly at the point of nullification. The extent of this   nullification should be visible in 
the quality of revival of the Hofstadter butterfly. The  method  that we use in our work is the numerical mean field approximation of 
unrestricted Hartree Fock. This method has the strength of handling both the electrical and magnetic 
sectors in the same footing(mean field) and being  able to handle spatial correlations in a self consistent fashion. 
\indent
It is a well known fact that the effect of disorder as well as interaction crucially depends on the lattice connectivity. If the lattice
connectivity or the effective physical dimension is larger, the effect of disorder or interaction is weaker. To explore 
this aspect we have considered in our study square as well as honeycomb lattice.  Recently it has been established that entanglement  is
an important aspect of a many body system which can reveal many salient  characteristics   very effectively, for example,
phase transition \cite{amico-2008}, topological order \cite{li-haldane}, edge states \cite{saptarshi-2016} etc. Previous
studies of entanglement on Hofstadter problem was mainly confined to bilayer systems \cite{schliemann,arovas-2012} or for cylindrical geometry \cite{moradi-2016}
where entanglement in momentum space was studied after tracing out one layer of physical system or one spatial direction respectively. In these cases
it was possible to obtain an analytical expression for entanglement entropy in momentum space. In our study we are investigating the usual entanglement in
physical lattice by integrating out a certain region  to explore how the spatial entanglement does evolve with magnetic field. Specifically
we see how the area law is being affected. We also  examine entanglement spectrum for the largest subsystem (which in our case is half the original system)
and observe useful patterns similar to Hofstadter pattern. The interplay of interaction and disorder in the entanglement entropy and entanglement spectrum is investigated thoroughly.

\indent
Our plan of presentation is following. In Sec. \ref{model}, we discuss in detail the main theoretical construction and approximation used.
This includes the derivation of mean-field approximation used in the paper as well as the description of Rudin-Shapiro disorder being implemented.
In Sec. \ref{subsys}, we describe the bi-partition of the square lattice and honeycomb lattice to define the reduce density matrix and hence
to calculate the entanglement entropy and entanglement spectra. In Sec. \ref{crit}, we provide our estimation of critical strength of interaction 
where the gap at half filling due to interaction and disorder nullify each other for square and honeycomb lattice. Followed by this, we explain 
our main results. In Sec. \ref{sec:spectra} we describe the effect of interaction and disorder on Hofstadter spectra for square and honeycomb lattice.
In Sec. \ref{sec:entangle} we describe the effect of interaction and disorder on entanglement entropy  and entanglement spectra.  Finally we conclude our results in Sec. \ref{discussion} with a discussion and put our finding in a larger perspective.

\section{Model and Method}
\label{model}
We now present and discuss the model Hamiltonian and scheme of dealing with interaction and disorder. We   implement a 
tight-binding single-band Hubbard model on square and honeycomb lattice. The reason to study both square and honeycomb lattice is to have a comparative study of two different two dimensional lattices with different  co-ordination number. Each site of a square lattice has four nearest neighbors and
for honeycomb lattice there are three nearest neighbors (effective lower dimension). As a result localization effect of disorder will be more prominent 
in honeycomb lattice and hence we expect a better recovery of Hofstadter butterfly in square lattice compared to honeycomb on switching 
on electronic interaction. We choose the Landau gauge $(0, Bx, 0)$ such that the components of the vector potential are $A_{x}=A_{z}=0, 
A_{y}=Bx$  to get uniform magnetic field along z-direction.  Therefore for the square lattice the phase appears only in the hopping along $y$-direction. The Hamiltonian for the square as well as honeycomb lattice  can be written as,
\begin{eqnarray} 
\label{main-h}
\mathcal{H}&=&-\sum_{i, \alpha, \sigma}(t_{i,\alpha}c^\dagger_{i\sigma}c^{}_{i+\delta_\alpha,\sigma}+h.c) -\sum_{i\sigma}(\epsilon_{i}-\mu)c^\dagger_{i\sigma}c^{}_{i\sigma} \nonumber \\
&&+\sum_{i}U n_{i\uparrow} n_{i\downarrow} .
\end{eqnarray}
 Here the label `$i$' represents a site of the two-dimensional square lattice or honeycomb lattice. The operator $c^\dagger_{i\sigma}$ ($c^{}_{i\sigma}$) creates (destroys) an electron of spin $\sigma$ at site `$i$'. We set the hopping `$t$' to be nearest neighbor only. We take a gauge such that $t_{i,y}= t e^{-i\phi_{i}},~ t_{i,x}=t, \alpha=x,y, \sigma=\uparrow, \downarrow$ where $\phi$ arises due to external magnetic field. $\delta_{\alpha}$ denotes nearest neighbor along $\alpha$ direction. The last term is the on-site Hubbard interaction term.  The chemical potential $\mu$ is calculated by demanding that there be exactly N/2 electrons in the problem. This is done by taking the average of the N/2-th and the (N/2 + 1)-th energy level. We set $t = 1$ uniformly for all the bonds in both the lattices. We make no further assumption about the magnetic order if any, and thus retain all spin indices in the formulas below.  \\
 
 To make progress with the Hamiltonian  given in \ref{main-h}, the last term i.e the Hubbard interaction is treated under Unrestricted Hartree-Fock(UHF) approximation ~\cite{claveau-2014,gupta1,gupta2}. To put it more clearly, the Hubbard onsite term in UHF is approximated
 by putting the product of onsite
fluctuations of the up and down spin electrons to zero.
\begin{eqnarray}
&&U(n_{i\uparrow}- \langle n_{\uparrow} \rangle)(n_{i\downarrow}- \langle n_{i\downarrow} \rangle)=0, or \nonumber \\ 
&&Un_{i\uparrow}n_{i\downarrow} = U(n_{i\uparrow} \langle n_{\downarrow} \rangle + \langle n_{i\uparrow} \rangle n_{i\downarrow} - \langle n_{i\uparrow} \rangle \langle n_{i\downarrow}\rangle)
\end{eqnarray}
 We will drop the last term because it does not contribute to the equation
 of motion. Once this approximation is
 implemented for the Hubbard interaction term, the on-site potential is modified as:
\begin{eqnarray}
&&\epsilon_{i\uparrow}^{'}=\epsilon_{i\uparrow}+U \langle n_{i\downarrow} \rangle,~\epsilon_{i\downarrow}^{'}=\epsilon_{i\downarrow}+U \langle n_{i\uparrow} \rangle 
\end{eqnarray}
And the corresponding decoupled Hamiltonian can be written as sum of 
up and down components as follows.

\begin{eqnarray} 
\label{MFH}
H_{mf}&&=\sum_{i}\epsilon_{i\uparrow}^{'}n_{i\uparrow}-
\sum_{i, \alpha}t_{i,\alpha}c^\dagger_{i\uparrow}c^{}_{i+\delta_\alpha,\uparrow} \nonumber \\
&&+\sum_{i}\epsilon_{i\downarrow}^{'}n_{i\downarrow}-\sum_{i, \alpha}t_{i,\alpha}c^\dagger_{i\downarrow}c^{}_{i+\delta_\alpha,\downarrow}\\
H_{mf}&&=H_{i\uparrow}+H_{i\downarrow} 
\end{eqnarray}
After writing the Hamiltonian as a sum of up and down components we start with 
an initial guess for $n_{i\uparrow}$ and $n_{i\downarrow}$. The Hamiltonian is
then diagonalized for both up and down spin electrons and new values of
$n_{i\uparrow}$ and $n_{i\downarrow}$ are generated. The process is iterated
until we reach a self consistent solution. Then one can calculate all the
physical quantities.  
Having discussed the mean-field approximation employed for Hubbard interaction we briefly describe the implementation of Peierls phase factor \cite{peierls-1933} due to magnetic field. We have taken the Landau gauge in which magnetic vector potential is represented as $\vec{A}= B x \hat{y}$
where $\hat{y}$ represents unit vector along $y$-direction. It is evident that for square lattice all the hopping parameter $t$ along $x$-direction is unchanged but the hopping elements along $y$-direction is
given by $t  \rightarrow t e^{i\phi_i}$ where $\phi_{i}= \int{A.dl}= \int{A_{y}dy} =Bx_{i}$, $x_{i}$ is the horizontal co-ordinate of $i$'th site as shown in Fig. \ref{fig:ent-subsystem}. However  for the honeycomb lattice, in addition to the hopping along $y$-direction, the hopping along the zig-zag chain extending along $x$ directions  would also acquire a phase because of the finite component along the $y$-direction. We denote   this phase as $\phi^{\prime}$ where $\phi^{\prime}= B x_i dy_i$. Now the translation vector along  the zig-zag chain is taken as $\vec{a}_{\pm}= \frac{\sqrt{3}}{2}  \hat{x}  \pm \frac{1}{2} \hat{y}$ which yields $dy= \pm dx/{\sqrt{3}}$ (in Fig. \ref{fig:ent-subsystem}, right panel, they are shown by light blue and pink thick lines). An elementary calculation shows that  $\phi^{'}_{i}= \int Bx_{i}dy_{i}= \int Bx_{i} dx_{i}(1/\sqrt{3})=  (1/\sqrt{3})B(x^{2}_{i+1}/2 - x^{2}_{i}/2)$ for both type of slanted bonds.   With the above convention it is easy to check that for square lattice the flux per plaquette  is invariant upon $B \rightarrow B+ 2 \pi$ and for the honeycomb lattice the corresponding shift is $B \rightarrow B + 2\pi/\sqrt{3} \approx B + 3.64$. As the Hofstadter spectra only depends on the flux per plaquette, the periodicity of Hofstadter spectra for square and honeycomb lattice is $2 \pi$ and $2\pi/\sqrt{3} \approx 3.64$ respectively. We note that here the strength of magnetic field is assumed to be in the unit of fundamental magnetic flux quanta.  In our numerical study we solve Eq. \ref{MFH} self-consistently using the above phase factors for the hopping parameters. We begin with a initial distributions of  $ \langle n_{i\sigma} \rangle$ (for a given realization of $\epsilon_i$) and diagonalize the effective single particle Hamiltonian as given in Eq. \ref{MFH} to obtain the new distribution of $\langle n^{\prime}_{i\sigma} \rangle$. The resulting distribution of $\langle n^{\prime}_{i\sigma} \rangle$s are taken as the initial values in the next iteration and the iteration continues until the initial distribution $\langle n_{i\sigma} \rangle$ and $\langle n^{\prime}_{i\sigma}\rangle$ converge within a prescribed accuracy.\\

\indent
The value of the site potentials $\epsilon_{i}$'s are assigned by the four letter Rudin-Shapiro sequence, which is wrapped around the lattice. Practically we implement this by an  one-dimensional representation of the lattice  as shown in Fig. \ref{fig:ent-subsystem}. The sequence is generated by using four letters $A, B, C, D$, the generating scheme is $A\rightarrow AB$, $B\rightarrow AC$, $C \rightarrow DB $, 
$D \rightarrow DC$ where $A, B, C, D$ represents the sequence for a given successive four sites along the one-dimensional array and $AB, AC, DB, DC$ represent the sequence for next four sites. The Rudin-Shapiro sequence \cite{narad-2015} therefore grows as $\it{ ABACABDBABACDCACABACABDBDCDBABDB...}$ and so on. In the present study  we have taken $A=0, B=1, C=0.5, D=1.5$ and this sequence is termed as $d=1$ throughout the manuscript.  In this study we have taken the Rudin-Shapiro sequence to simulate the disorder in the system for simplicity. To take the effect of disorder into account properly, one needs to take a configuration average of many realization of disorder which is computationally very heavy specially in the presence of interaction. However we think that the result obtained within the Rudin-Shapiro sequence generated disorder is able to produce the qualitatively identical  results as obtained from proper configuration average.

\section{Details of subsystem for entanglement calculation}
\label{subsys}
\begin{figure}[!htb]
\centering
  \includegraphics[width=.99\linewidth]{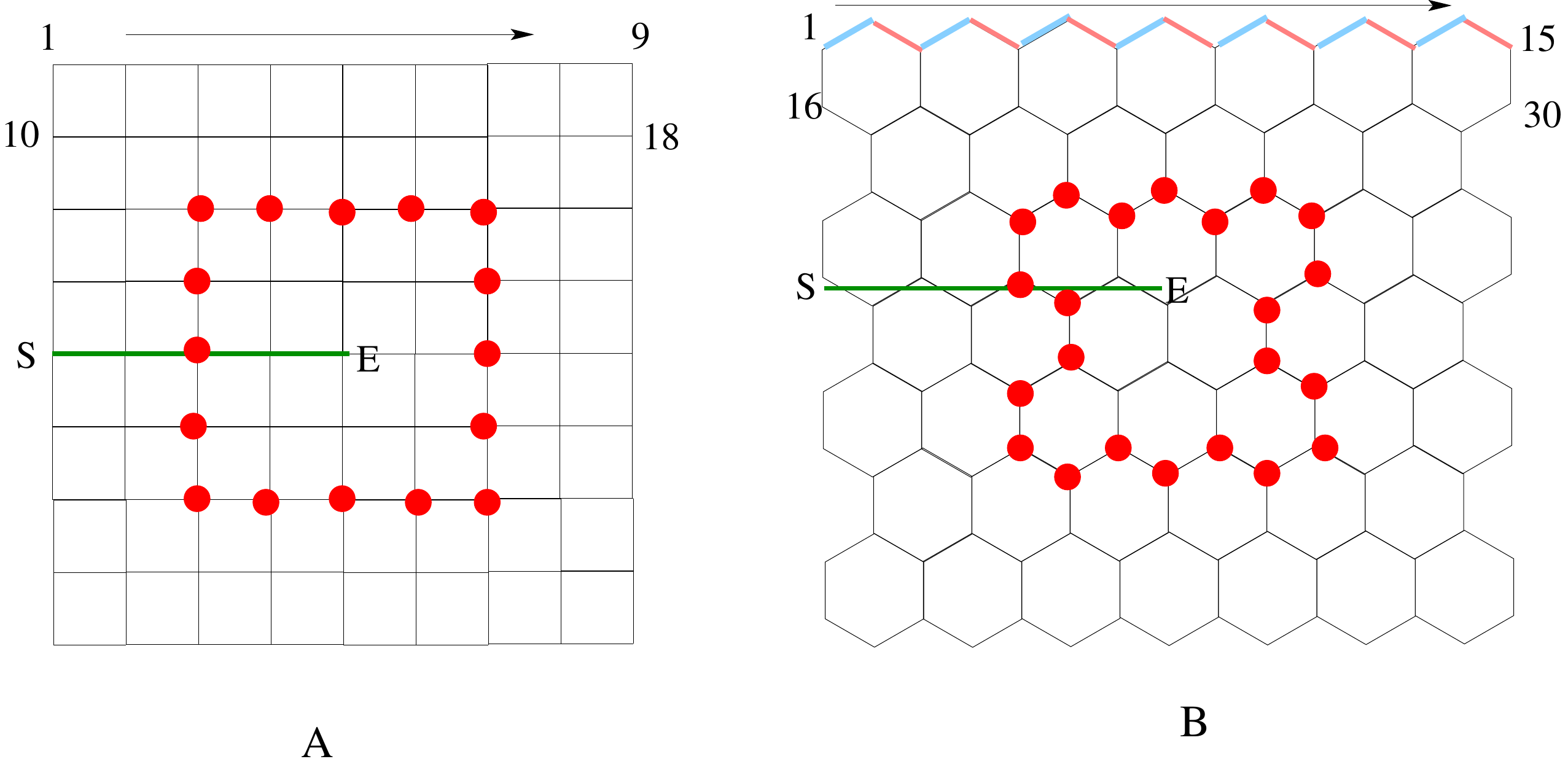}
  \caption{Detail geometry of   square and honeycomb lattice used for studying the Hofstadter effect. The numeric attached to the figures show the way two dimensional lattice is mapped to one dimensional array for applying Rudin-shapiro sequence. The region marked by red circles in its boundary denotes the geometry of an elementary subsystem for evaluation of entanglement entropy. For details see the text in Sec. \ref{subsys}.}
  \label{fig:ent-subsystem}
\end{figure}

Having discussed our theoretical model and approximation scheme in detail, we briefly describe our scheme of entanglement estimation in the present study. As described earlier  we have taken a square lattice of dimension $42 \times 42$ with an open boundary condition. To estimate the 
entanglement entropy we define  various subsystem as follows. We take a square of dimension $ n \times n$ and increase `$n$' from 3 to 21 implying that we take
a subsystem whose size includes $m \times m$ squares with `$m$' taking values 2 to 20. In Fig. \ref{fig:ent-subsystem} A, we have shown such a subsystem
having dimension of $ 4 \times 4$ squares. The entanglement spectrum has been investigated only for the largest square subsystem when $m=21$.
The entanglement entropy has been calculated by usual procedure of diagonalizing the correlation matrix of the subsystem~\cite{ent} for free fermionic system. In all the figure entanglement entropy has been plotted  for various subsystem represented by $m$ for the $m \times m$ squares. For the honeycomb lattice, we have presented a cartoon picture in Fig. \ref{fig:ent-subsystem} B. The various subsystem has $m$ zig-zag chains where in each chain has $m+1$ sites. In the Fig.  
\ref{fig:ent-subsystem} B, we have represented the upper row of such subsystems by red filled circles and the region bounded by the red filled circles shows a generic subsystem
of $4 \times 5$ sites. The largest value of $m$ taken by us is 23 and entanglement spectrum is investigated for this largest subsystem only. Entanglement spectrum is defined as  the eigenvalues of the reduced density matrix. The eigenvalues of the reduced density matrix is related to the eigenvalues of the correlation function matrix as shown in \cite{ent}.  If $\lambda_k$'s are the eigenvalues of the correlation matrix then one can obtain the eigenvalues of the reduced density matrix. In appendix we elucidate this relation. This intern enables one to calculate the von-Neuman entanglement entropy of a non-interacting system completely determined by the formula \cite{amico-2008,eisert-2010},

\begin{eqnarray}
\mathcal{S}_{en}&&= \sum_{k} -\left(\lambda_k log \lambda_k+ (1-\lambda_k) log (1-\lambda_k)   \right)
\end{eqnarray}
 In our study we have evaluated the above quantity. Apart from this we have also plotted $\lambda_k$ for the largest subsystem for different magnetic field. The distribution of $\lambda_k$ is directly related with the physical spectrum and entanglement spectrum\cite{fidkowski-2010}.

\section{Critical values of interactions}
\label{crit}

We have already explained that our motivation was to investigate the extent of revival of the Hofstadter butterfly due to interplay of disorder and interaction. Thus the pertinent question is what is the order parameter that should be a good indicator of such interplay. It is known that an onsite Hubbard interaction opens a Mott-Hubbard gap at half filling. The same is true for the present study as well. For a self consistent study in the presence of interaction, the shape of the butterfly nature of the spectrum remains the same though a gap opens at half filling. In the presence of small and moderate disorder the fractal nature of the butterfly spectrum starts to melt away giving a structure less spectrum. In our study  there are interaction and disorder both. The gap at half filling at B = 0 is the physical indicator of how the competition between these two mechanisms manifest. The localization effect of disorder dominates till a critical value of onsite Hubbard interaction $U_c$ is reached beyond which a gap at half-filling opens up due to electronic correlation and the system becomes an insulator because of electron-electron interaction. There are doubly occupied sites below $U_c$ but above it the onsite interaction dominates and double occupancies become energetically unfavourable. Therefore $U_c$ is the point at which the two effects nullify each other.
\begin{figure}[!htb]
\centering
\includegraphics[width=.8\linewidth,height=5cm]{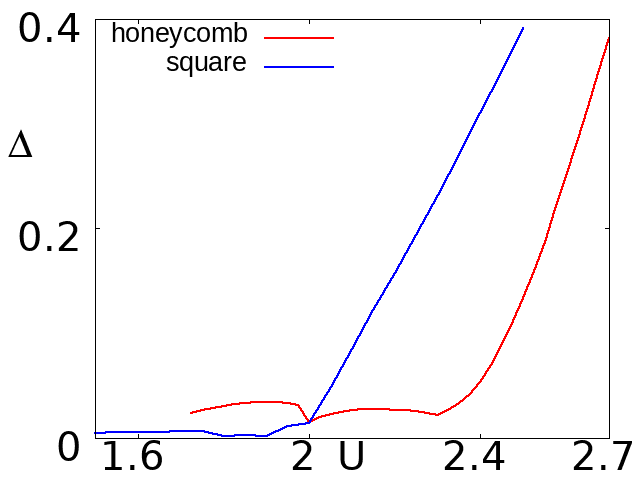}
  \caption{The gap at half filling,  $\Delta$, is shown for square and honecomb lattice in the presence of disorder.}
  \label{fig:gap}
\end{figure}
 For square lattice and  Honeycomb lattice we obtain the value of $U_{cr}$  as 1.9 and 2.3 respectively as shown in Fig. \ref{fig:gap}. The location of $U_{cr}$ as described is not very precise
due to finite size effect. We have found that for various system sizes  there is a non-uniformity in gap at half filling  and the location of $U_{cr}$ varies though they are very near to 1.9 and 2.3 for square and honeycomb lattice respectively. We have zeroed in at these values of $U_{cr}$ after a careful comparison of data for various system sizes such that finite size effect is minimal. For this reason we have chosen a system size of $42 \times 42$ for square lattice and $45 \times 46$ for honeycomb lattice.
\section{Results}
\begin{figure}[!htb]
\begin{subfigure}{.23\textwidth}
  \centering
  \includegraphics[width=.99\linewidth]{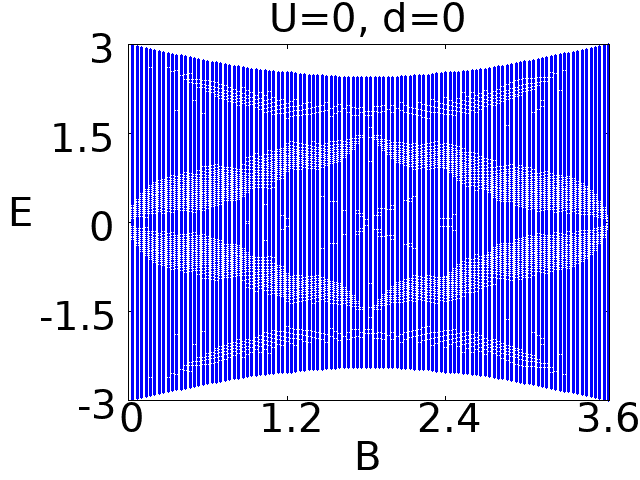}
  \label{hex1}
\end{subfigure}%
\begin{subfigure}{.23\textwidth}
  \centering
  \includegraphics[width=.99\linewidth]{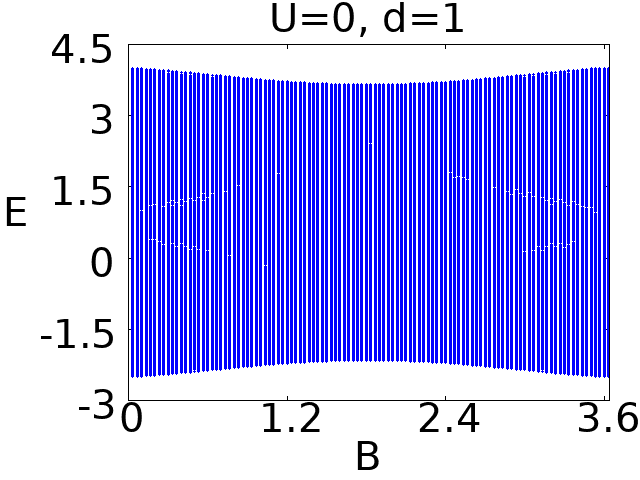}
  \label{hex4}
\end{subfigure}%

\begin{subfigure}{.23\textwidth}
  \centering
  \includegraphics[width=.99\linewidth]{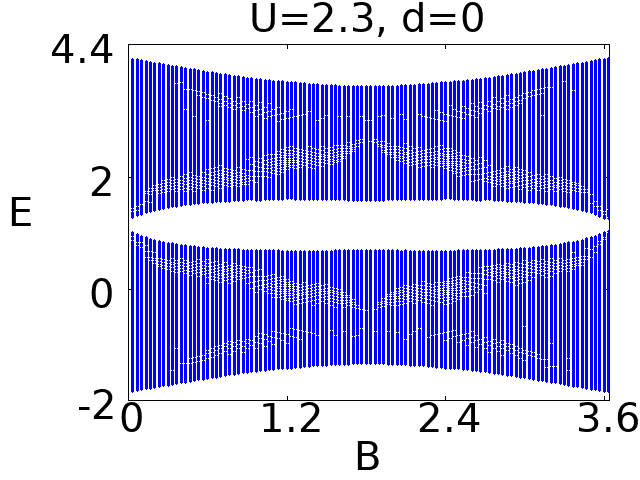}
  \label{hex7}
\end{subfigure}%
\begin{subfigure}{.23\textwidth}
  \centering
  \includegraphics[width=.99\linewidth]{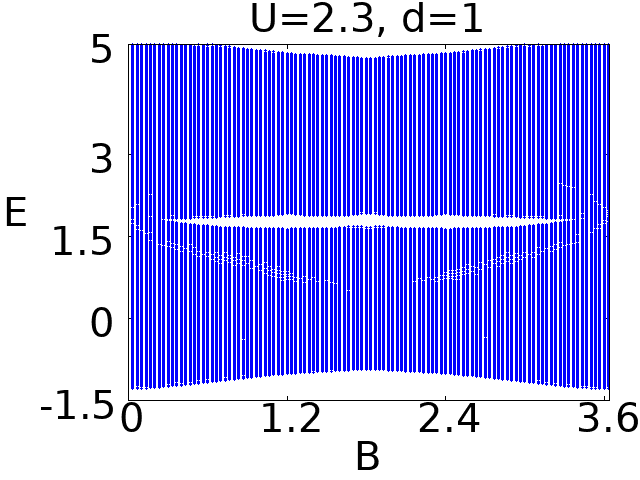}
  \label{hex10}
\end{subfigure}%

\caption{Energy spectrum(E) is plotted against magnetic field(B) in honeycomb lattice for various combinations of interaction and disorder. The dual effect of interaction and disorder is similar that has been seen in square lattice. However due to less lattice coordination number the gap due to interaction is not completely removed by the disorder.  }
\label{hcspectra}
\end{figure}

We now move on to describe our main results. From now on   `E' stands for the Hofstadter spectrum, `S' stands for entanglement
entropy for subsystem as plotted against different magnetic field and `T' denotes entanglement spectra for half subsystem as defined before.
In Fig. \ref{sqspectra} and Fig. \ref{hcspectra} we plotted the Hofstadter spectra for square and honeycomb lattice respectively. For entanglement entropy and spectrum for square lattice we refer Fig. \ref{sqentropy}  and Fig. \ref{sqcorre} respectively. Similarly for honeycomb lattice, we use Fig. \ref{hcentropy}  and Fig. \ref{hccorre} for entanglement entropy and entanglement spectra respectively. 
In all the above mentioned figures   the upper left panel represents the plot  in the absence of disorder and interaction, upper right panel represents the case when disorder  is present in the absence of interaction. In the lower left panel we present the plot when interaction is present but there is no disorder. Finally in the lower right panel, we present the result in the case of simultaneous presence of disorder and interaction. With these guidelines for understanding the figures we now move on to discuss the effect of interaction and disorder on Hofstadter spectra.
\subsection{Effect of disorder and interaction on spectrum}
\label{sec:spectra}
\subsubsection{Square lattice}
\label{squarelat}
\begin{figure}[!htb]
\begin{subfigure}{.23\textwidth}
  \centering
  \includegraphics[width=.99\linewidth]{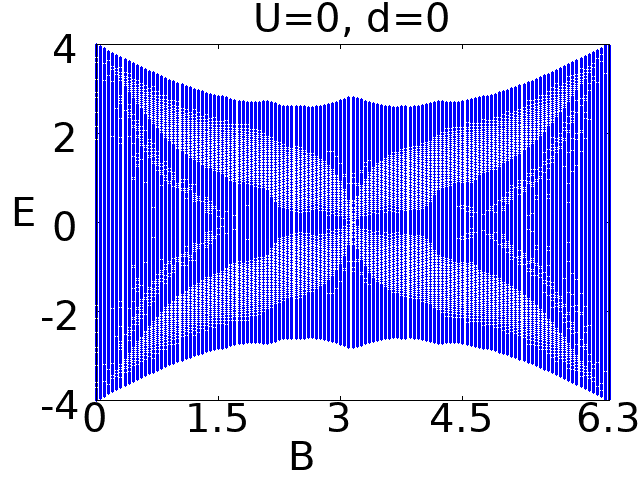}
  \label{sq1}
\end{subfigure}%
\begin{subfigure}{.23\textwidth}
  \centering 
  \includegraphics[width=.99\linewidth]{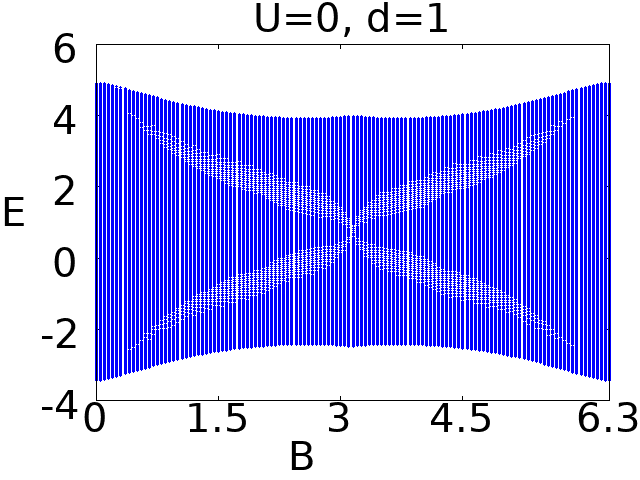}
  \label{sq4}
\end{subfigure}%

\begin{subfigure}{.23\textwidth}
  \centering
  \includegraphics[width=.99\linewidth]{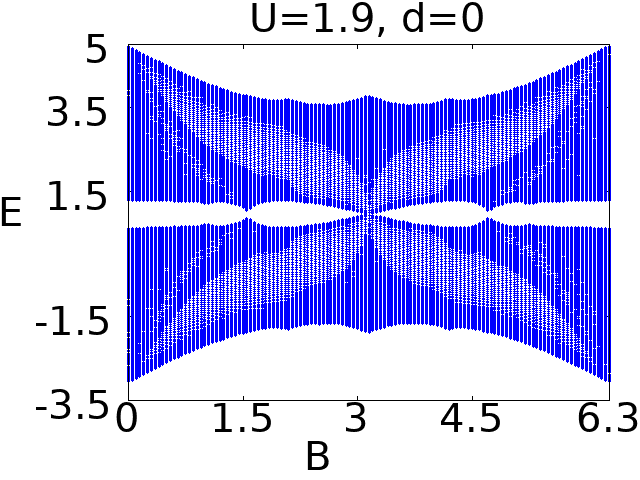}
  \label{sq7}
\end{subfigure}%
\begin{subfigure}{.23\textwidth}
  \centering
  \includegraphics[width=.99\linewidth]{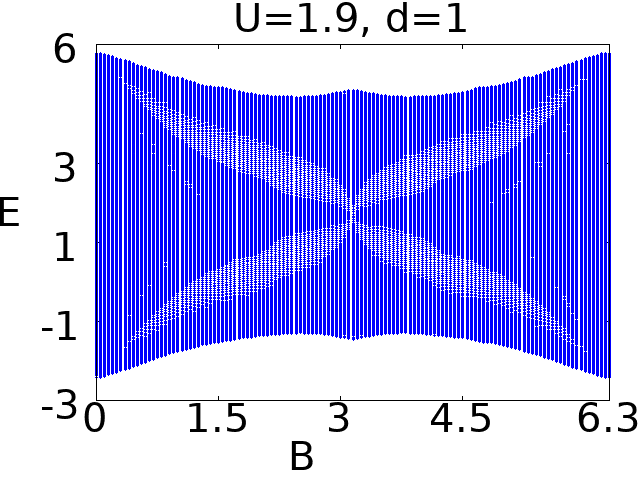}
  \label{sq10}
\end{subfigure}%

\caption{ Energy(E) spectrum is plotted with respect to magnetic field(B) in square lattice for various combinations of interaction and disorder as represented by the header of each plot. As can be seen disorder smears out the characteristic butterfly nature of the spectrum and interaction opens up a gap at half filling. The disorder in the presence of interaction successfully closes the gap and recovers the butterfly nature of the spectrum to some extent.}
\label{sqspectra}
\end{figure}

In Fig.\ref{sqspectra}, we have presented the effect of interaction and disorder graphically for square lattice. From the  convention of different parameters as mentioned earlier we observe from Fig. \ref{sqspectra} that as we turn on the disorder,   Hofstadter spectrum loses its fractal/butterfly  nature. The given disorder  inserts states into the gaps and is strong enough to smear out the fractal character and the spectrum almost looks like a continuous band though the lowest eigenvalue still retains the reflection symmetry with respect to $B=\pi$. 
Now we discuss the effect of interaction without disorder. Turning on interaction causes a gap in the Hofstadter spectrum retaining its overall  butterfly like structure as evident from lower left panel of Fig. \ref{sqspectra}.
Now as we turn on the disorder in the presence of interaction we observe the resurrection of Hofstadter butterfly for a critical value of interaction. This resurrection or revival  is marked by the appearance of the largest wing or gap in the spectra. Though we note that the width of the wing has decreased and the subdominant wings are also faintly present there. This is also accompanied with vanishing gap at half filling. Thus we find a clear evidence that the localization effect due to disorder and interaction play against each other and results in the revival of butterfly spectra to some extent.  Now we move on to explain the results for honeycomb lattice.

\subsubsection{Honeycomb single layer}
\label{hexalat}

For the Honeycomb lattice, we observe similar effect due to disorder and
 interaction in the Hofstadter spectrum. The disorder (generated by (0 0.5 1 1.5) Rudin
 Shapiro sequence) is strong enough to smear out  the entire Hofstadter
 butterfly structure almost completely. The effect of disorder is stronger than the square lattice which
can be attributed to reduced co-ordination number.  Again in the absence of disorder, the self consistent analysis of the Hubbard interaction gives a gap at half filling without otherwise disturbing the
 Hofstadter pattern. The induced Hubbard gap is larger in magnitude in comparison to square lattice. Further  for the square lattice the interaction introduced gap is minimum(maximum) at middle(end points) where as for honeycomb lattice it is opposite. We think this is due to the  Lieb's theorem \cite{lieb-1994} which states that for square lattice the minimum energy corresponds to $\pi$ flux per plaquette where as for honeycomb lattice the corresponding flux is zero. Furthermore in simultaneous presence of  disorder and interaction  we find that Hubbard gap is minimized
 and  small revival of the Hofstadter spectrum at $U=2.3$. However the revival of butterfly spectra is stronger in square lattice due to increase in coordination number. The present disorder is  strong enough so  that  the strength of $U$ required to nullify the effect of the same strength of disorder is also
 larger in comparison to square lattice due to the decrease in number of nearest neighbor hopping.

\subsection{Effect of interaction and disorder on Entanglement}
\label{sec:entangle}
\subsubsection{Square lattice}
\begin{figure}[!htb]
\begin{subfigure}{.23\textwidth}
  \centering
  \includegraphics[width=.99\linewidth]{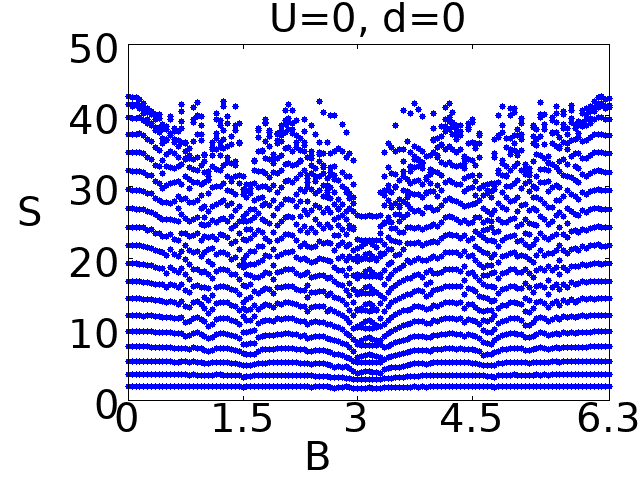}
  \label{sq2}
\end{subfigure}
\begin{subfigure}{.23\textwidth}
  \centering
  \includegraphics[width=.99\linewidth]{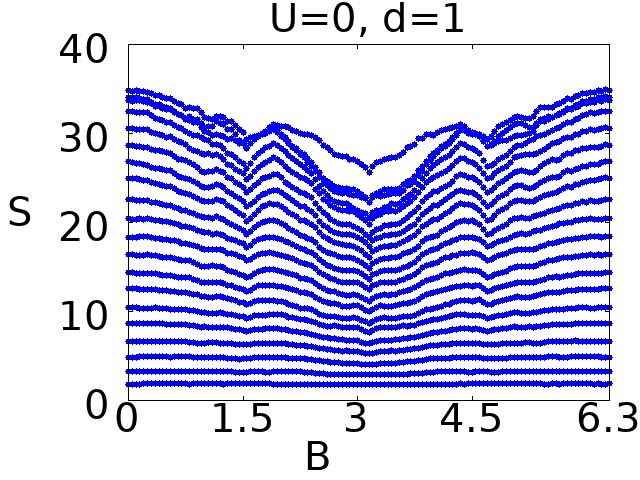}
  \label{sq5}
\end{subfigure}

\begin{subfigure}{.23\textwidth}
  \centering
  \includegraphics[width=.99\linewidth]{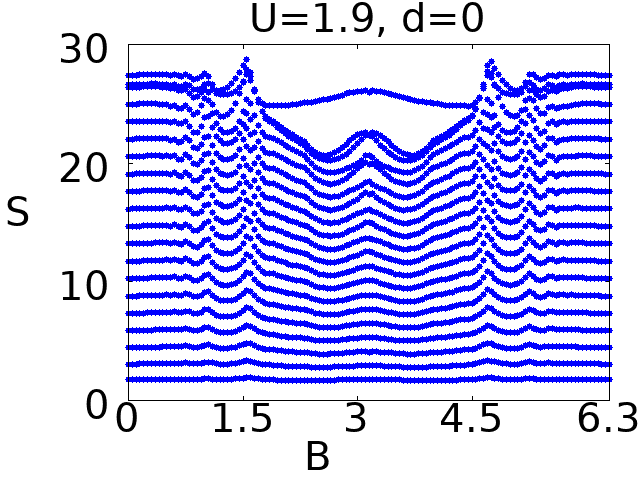}
  \label{sq8}
\end{subfigure}
\begin{subfigure}{.23\textwidth}
  \centering
  \includegraphics[width=.99\linewidth]{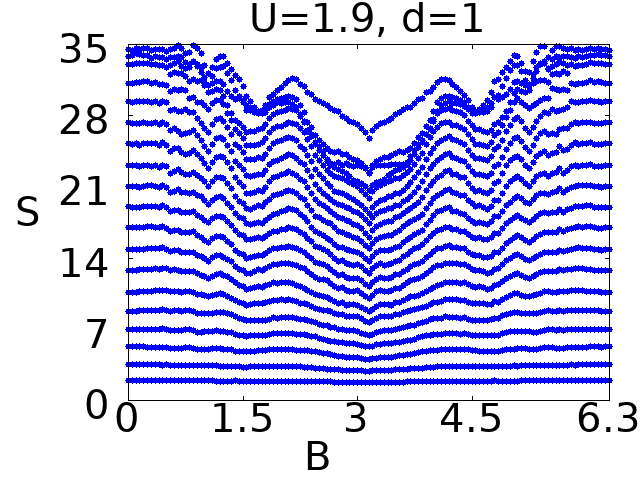}
  \label{sq11}
\end{subfigure}
\caption{ Entanglement entropy(S) is plotted ahainst magnetic field(B) in square lattice for various combinations of disorder and interactions. As one moves vertically for a given magnetic field, the different values of entanglement entropy corresponde to different block sizes starting from $4 \times 4$  to $L/2 \times L/2$. In general the quidistant entanglement entropy signify area law though there exist some departure as explained in the text.  }
\label{sqentropy}
\end{figure}

\begin{figure}[!htb]
\begin{subfigure}{.23\textwidth}
  \centering
  \includegraphics[width=.99\linewidth]{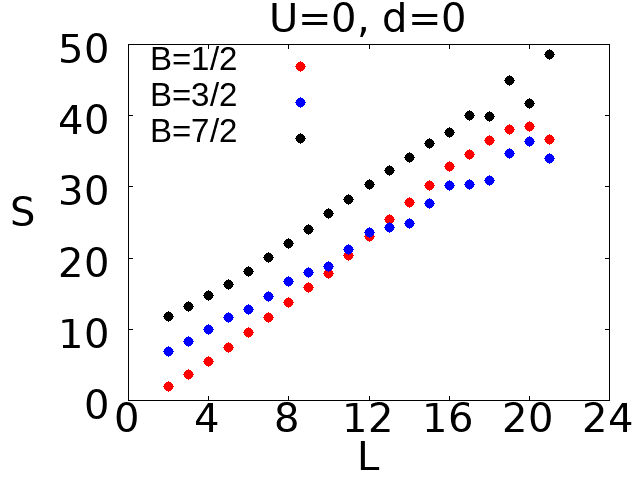}
  \label{sqarea}
\end{subfigure}
\begin{subfigure}{.23\textwidth}
   \centering
 \includegraphics[width=.99\linewidth]{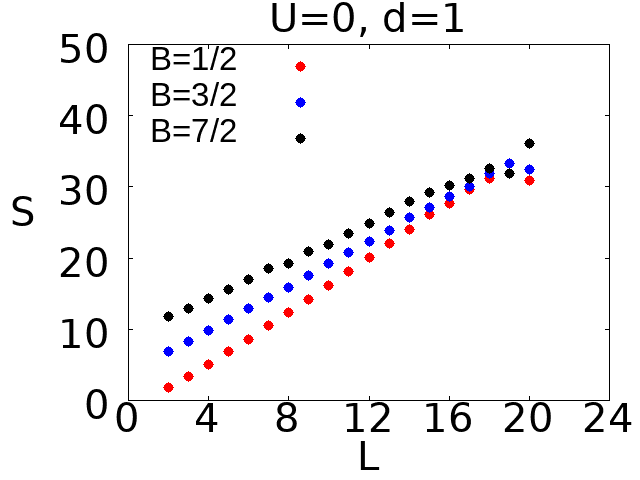}
\label{sqdarea}
\end{subfigure}
\begin{subfigure}{.23\textwidth}
  \centering
  \includegraphics[width=.99\linewidth]{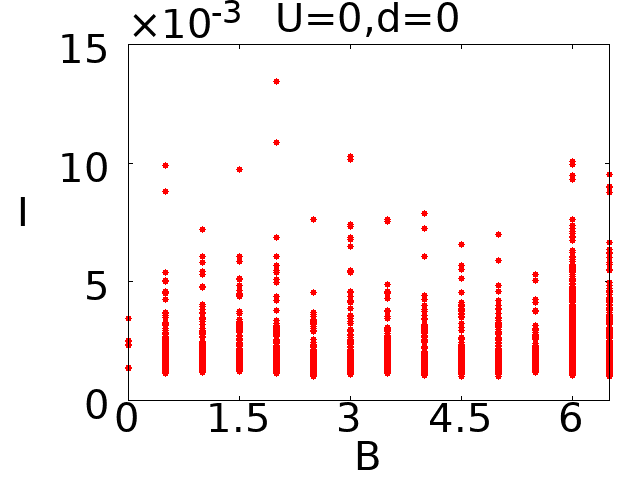}
  \label{sqipr}
\end{subfigure}
\begin{subfigure}{.23\textwidth}
   \centering
 \includegraphics[width=.99\linewidth]{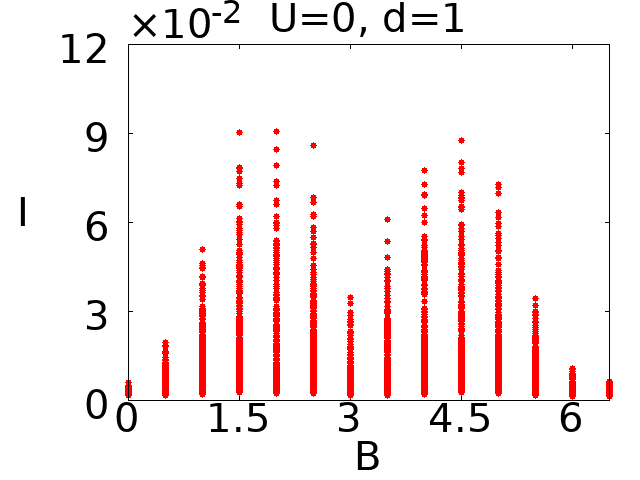}
  \label{sqdipr}
 \end{subfigure}
 \begin{subfigure}{.23\textwidth}
  \centering
  \includegraphics[width=.99\linewidth]{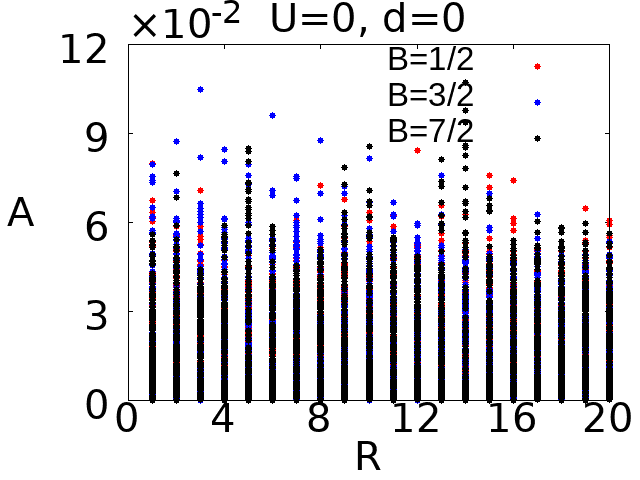}
  \label{sqedge}
\end{subfigure}
\begin{subfigure}{.23\textwidth}
   \centering
 \includegraphics[width=.99\linewidth]{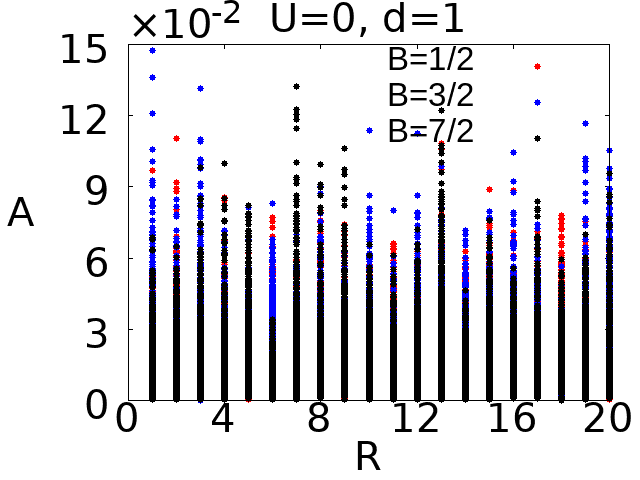}
\label{sqdedge}
\end{subfigure}
\caption{In the upper left panel entanglement entropy has been plotted  against subsystem sizes for representative values of magnetic field, the upper right panel shows the same in the presence of disorder. The middle left and right panel shows inverse participation ratios for various magnetic fields for uniform case and in the presence of disorder respectively.  In the lower left panel and right panel we show amplitude of wave function from edge to the center for uniform case and in the presence of disorder only.}
\label{sq-area-ipr-edge}
\end{figure}

\begin{figure}[!htb]
\begin{subfigure}{.23\textwidth}
  \centering
  \includegraphics[width=.99\linewidth]{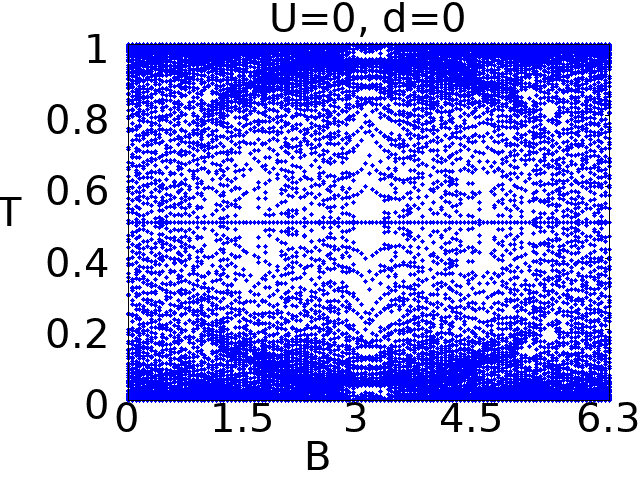}
  \label{sq3}
\end{subfigure}
\begin{subfigure}{.23\textwidth}
  \centering
  \includegraphics[width=.99\linewidth]{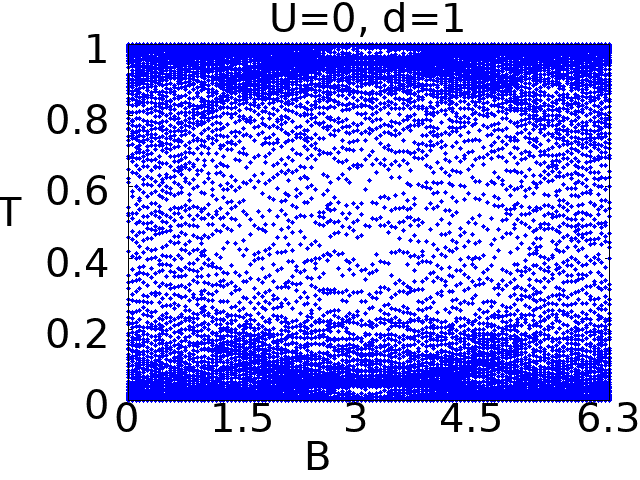}
  \label{sq6}
\end{subfigure}
\begin{subfigure}{.23\textwidth}
  \centering
  \includegraphics[width=.99\linewidth]{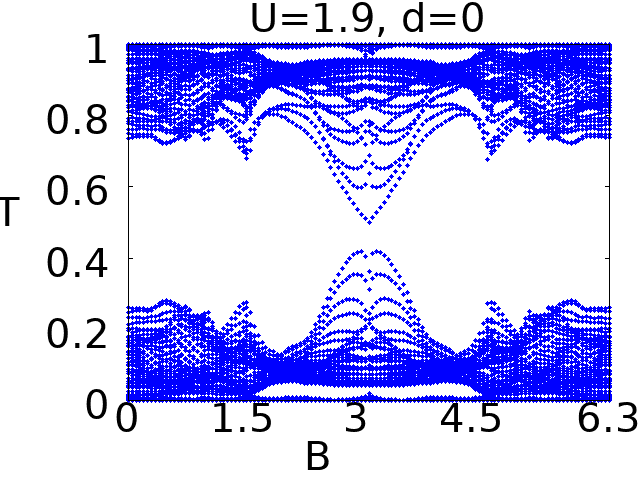}
  \label{sq9}
\end{subfigure}
\begin{subfigure}{.23\textwidth}
  \centering
  \includegraphics[width=.99\linewidth]{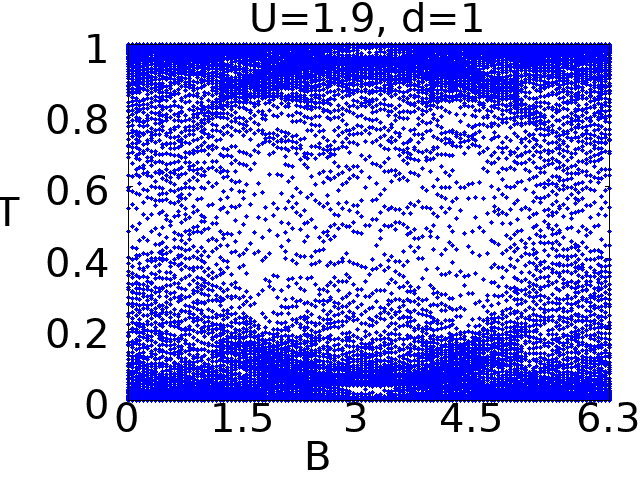}
  \label{sq12}
\end{subfigure}
\caption{The correlation function spectrum (T) is plotted for largest subsystem in square lattice. The headers of each plot denote the combinations of disorder and interactions as before. In the absence of disorder and interaction the spectrum does have a reflection symmetry. The interaction opens up a gap and disorder in general smears out the spectrum devoid of any special feature.}
\label{sqcorre}
\end{figure}

Having described the effect of interaction and disorder on Hofstadter spectrum we now discuss their effect on  entanglement. We have investigated entanglement entropy for various subsystem sizes as described in Sec. \ref{subsys} and also examined the correlation matrix spectrum for the largest subsystem whose dimension is half the full system. We begin our discussion with square lattice. In Fig. \ref{sqentropy}, we have plotted the entanglement entropy for various situations( interaction and disorder as mentioned earlier).  In each of these plot the horizontal axis represents magnetic field and  vertical axis denotes entanglement entropy for various block sizes as described in Fig. \ref{fig:ent-subsystem} and in Sec. \ref{subsys}. This imply that for a given magnetic field say at $B=0$, the fist blue point vertically denotes entanglement entropy of block size $2 \times 2$, the second points correspond to block size $4 \times 4$ and so on.   Thus an equispaced entanglement entropy for a given magnetic field ensures the area law to be maintained. From the Fig. \ref{sqentropy} we observe that for uniform case area law is better maintained for  magnetic fields close to zero or $2 \pi$. For magnetic field near $\pi$ area law is not maintained well for uniform case. However area law is shown to improve in the presence of disorder and interaction and both as we find the entanglement entropy to be more equispaced.  For better understanding, in Fig. \ref{sq-area-ipr-edge} upper left and right panel we have shown the entanglement entropy versus the length (L) (length of one side of square subsystem) for three representative magnetic field for uniform case and in the presence of disorder respectively. For better  representation entanglement entropy for higher magnetic field are shifted upward by 7.5 and 15 respectively. It is clear that the disorder induces better area law. However there seems to be a departure from area law for larger subsystems. To understand this we have plotted in the middle left and right panel the inverse participation ratio (see Appendix for definition) for uniform case and in the presence of disorder respectively at half filling.   It is clear that for the uniform case, the small values of IPR signify extended wave functions and the area  law is not expected to be maintained for this. For disordered case large values of IPR signify localization however for magnetic field near $\pi$, small values of IPR denotes delocalized or quasi-delocalized wave function which might cause such violation of area law for larger subsystem. In the lower left and right panel of Fig. \ref{sq-area-ipr-edge}, we plotted amplitude of wave functions from boundary to the center and we always find a significant probability of some wave functions denoting quasi-delocalized nature of the wave functions. As the entanglement entropy is a contribution from all the eigenfunctions up to half filling, it is rather difficult to come at definite conclusion for such area law violation. At most it points out to inhomogenous and extended nature of wave functions.  In appendix we have shown more elaborately how the IPR and amplitude of wave function varies for various system sizes and for various strength of interaction and disorder. This shows that there is always some quasi delocalized wave functions whose contributions to the entanglement entropy depend on the vary nature of the wave functions and the subsystem.  It may be noted that  such delocalized wave function has been experimentally observed in disordered Chern insulators \cite{udvas-2021}. Interestingly  entanglement entropy shows a local minima at B=$\pi$ for disordered case but shows a local maxima in the presence of interaction. To understand the entanglement entropy and its behavior near $B=\pi$, we plot the entanglement spectra for the largest subsystem as shown in Fig. \ref{sqcorre}. We observe that the gap in the entanglement spectrum follows directly from the gap at half filling shown in \ref{sqspectra} conforming the correspondence between entanglement spectra and physical spectra \cite{fidkowski-2010}. For example in Fig. \ref{sqcorre} lower left panel, we find that the electronic  interaction opens a gap at half filling but the magnitude of the gap depends on the magnetic field. At certain magnetic field, the gap
is minimum and at those values of magnetic fields we find that entanglement entropy depart from the area law. Also at these  values of magnetic field entanglement spectrum shows a  reduced gap. If we look at the entanglement spectrum for the pure case, we do note that there are some fractal structure as well. The magnetic  field length scale also present itself an important length scale which competes with the localization effect of interaction and at certain magnetic field, the effect of interaction is reduced to a minimum. This may cause enhancement of entanglement entropy and reduced gap in entanglement spectrum.

\subsubsection{Honeycomb lattice}

\begin{figure}[!htb]
\begin{subfigure}{.23\textwidth}
  \centering
  \includegraphics[width=.99\linewidth]{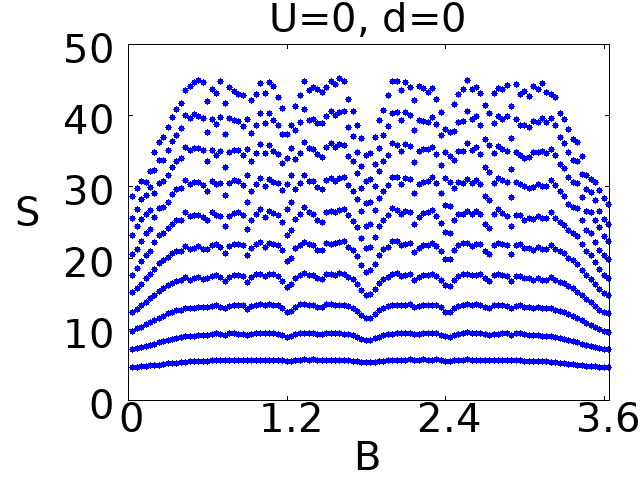}
  \label{hex2}
\end{subfigure}
\begin{subfigure}{.23\textwidth}
  \centering
  \includegraphics[width=.99\linewidth]{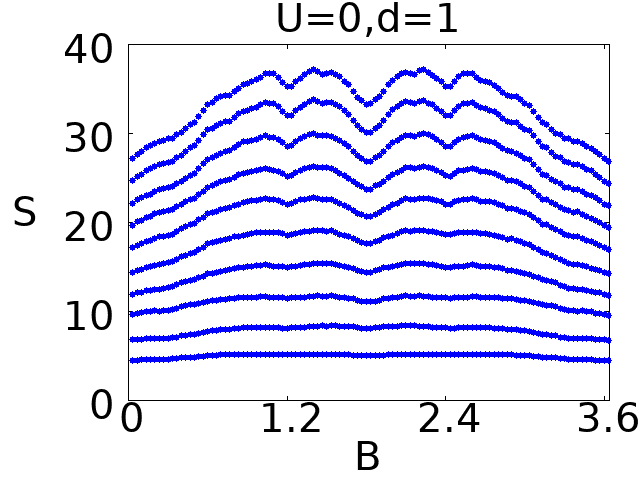}
  \label{hex5}
\end{subfigure}
\begin{subfigure}{.23\textwidth}
  \centering
  \includegraphics[width=.99\linewidth]{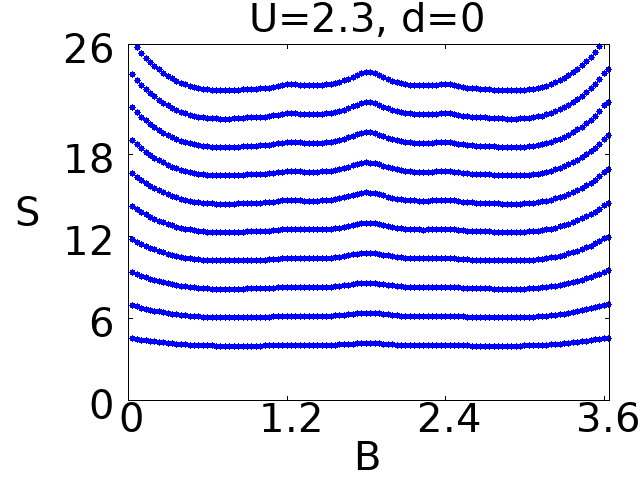}
  \label{hex8}
\end{subfigure}
\begin{subfigure}{.23\textwidth}
  \centering
  \includegraphics[width=.99\linewidth]{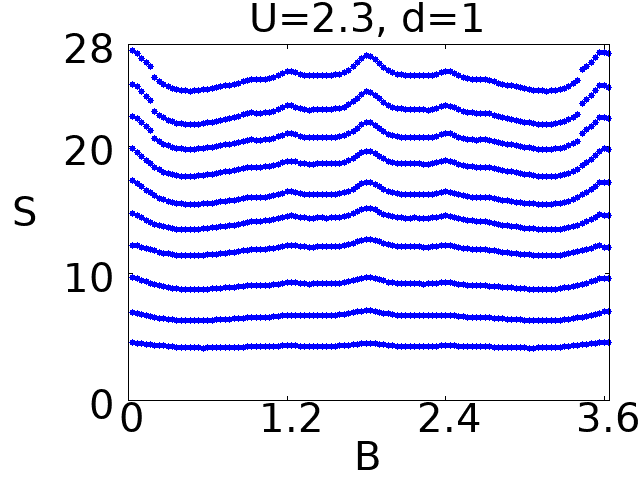}
  \label{hex11}
\end{subfigure}
\caption{Entanglement entropy (S) is plotted against magnetic field(B) in honeycomb lattice for various values of disorder and interactions. Various values of  entanglemenet entropy for a given magnetic field corresponde to different block sizes. The blocks are taken from $4 \times 5$  to $L/2 \times (L/2+1)$ with successive increment of 2 sites in each directions. The equal spacings between successive values of entanglement entropy at a given magnetic field signify strict area law maintained.}
\label{hcentropy}
\end{figure}

\begin{figure}[!htb]
\begin{subfigure}{.23\textwidth}
  \centering
  \includegraphics[width=.99\linewidth]{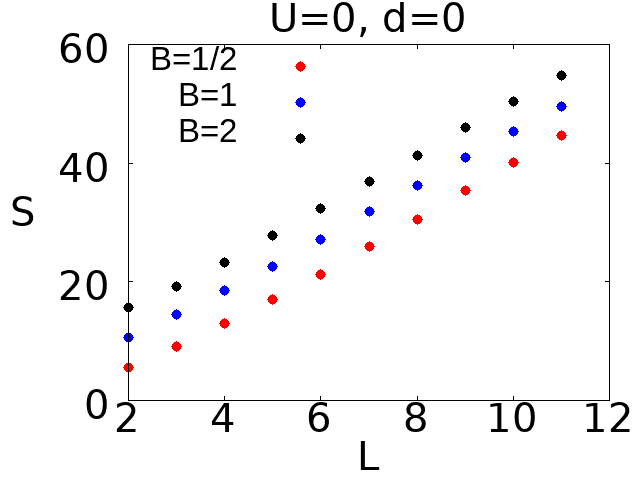}
  \label{hc-area1}
\end{subfigure}
\begin{subfigure}{.23\textwidth}
   \centering
 \includegraphics[width=.99\linewidth]{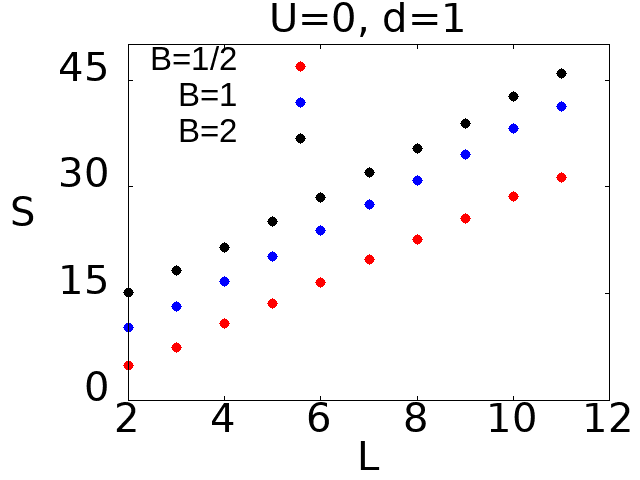}
\label{dhc-area}
\end{subfigure}
\begin{subfigure}{.23\textwidth}
  \centering
  \includegraphics[width=.99\linewidth]{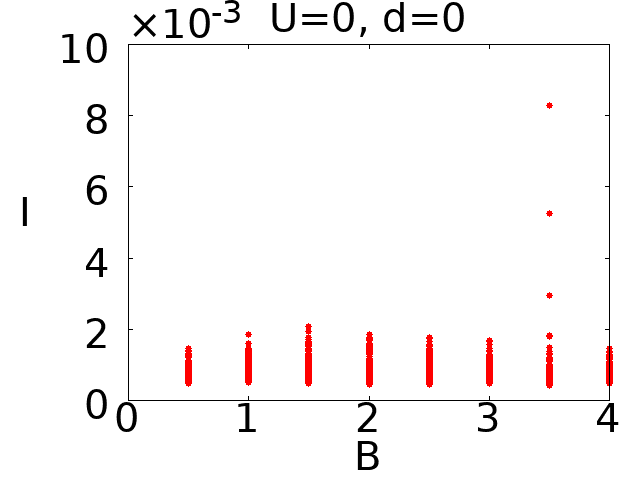}
  \label{hc-ipr}
\end{subfigure}
\begin{subfigure}{.23\textwidth}
   \centering
 \includegraphics[width=.99\linewidth]{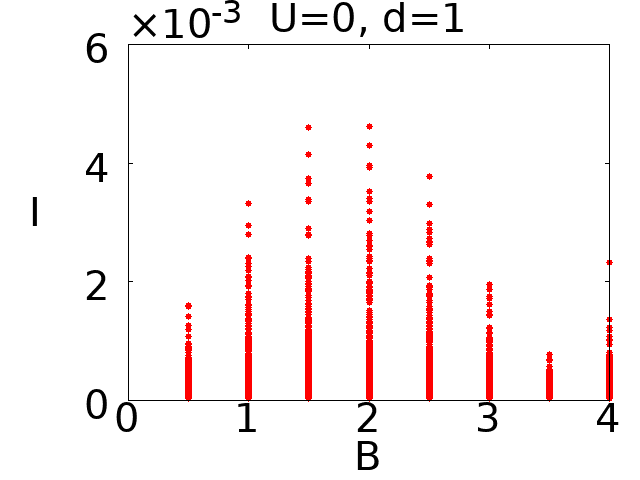}
  \label{hc-d-ipr}
 \end{subfigure}
 \begin{subfigure}{.23\textwidth}
  \centering
  \includegraphics[width=.99\linewidth]{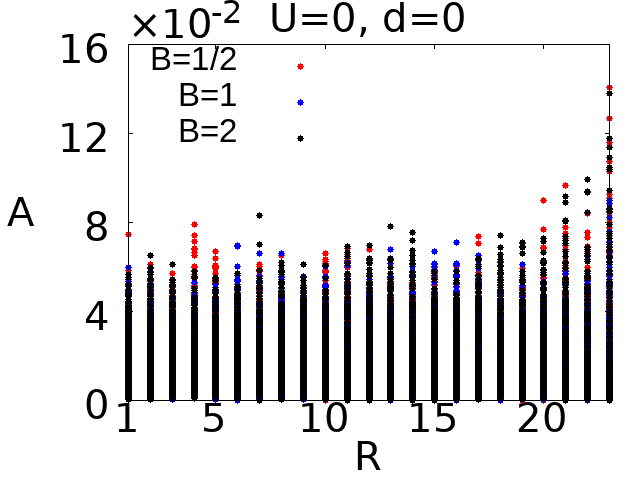}
  \label{hc-edge}
\end{subfigure}
\begin{subfigure}{.23\textwidth}
   \centering
 \includegraphics[width=.99\linewidth]{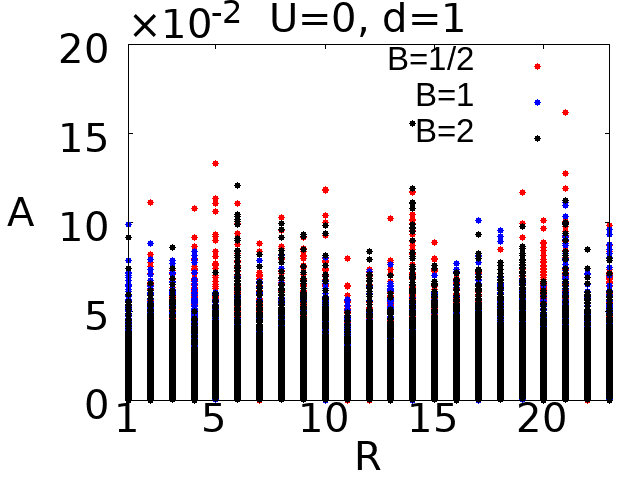}
\label{hc-d-edge}
\end{subfigure}
\caption{In the upper left panel entanglement entropy has been plotted  against subsystem sizes for representative values of magnetic field, the upper right panel shows the same in the presence of disorder. The middle left and right panel shows inverse participation ratios for various magnetic fields for uniform case and in the presence of disorder respectively.  In the lower left panel and right panel we show amplitude of wave function from edge to the center for uniform case and in the presence of disorder only.}
\label{hc-area-ipr-edge}
\end{figure}

\begin{figure}[!htb]
\begin{subfigure}{.23\textwidth}
  \centering
  \includegraphics[width=.999\linewidth]{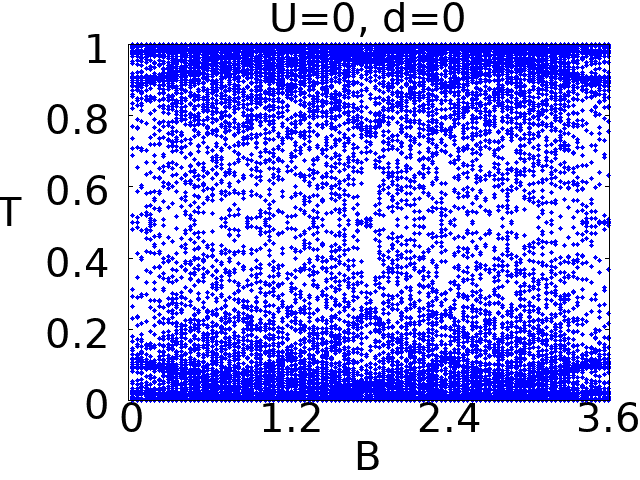}
  \label{hex3}
\end{subfigure}
\begin{subfigure}{.23\textwidth}
  \centering
  \includegraphics[width=.999\linewidth]{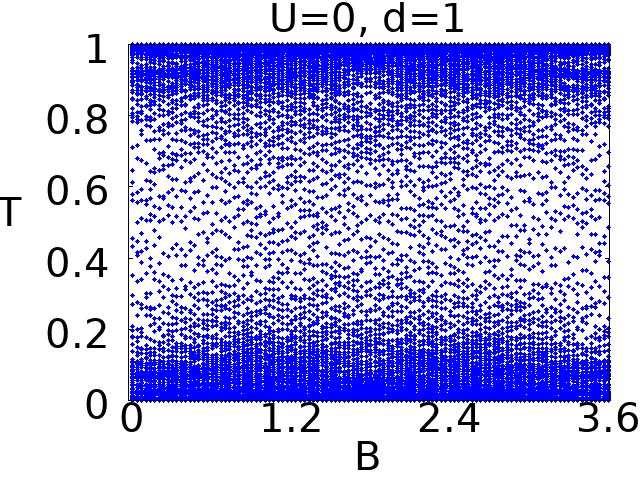}
  \label{hex6}
\end{subfigure}
\begin{subfigure}{.23\textwidth}
  \centering
  \includegraphics[width=.999\linewidth]{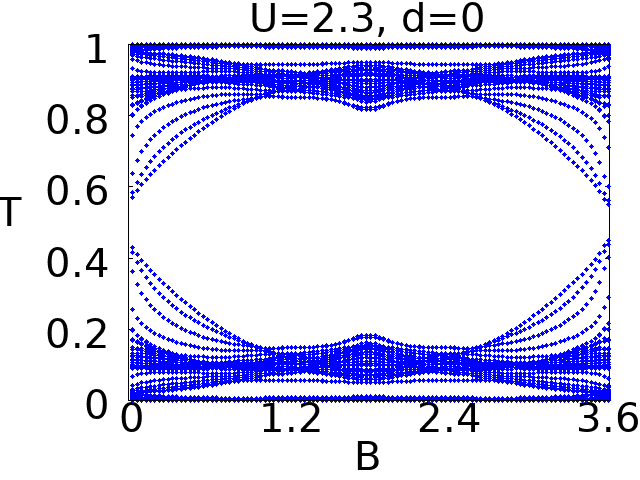}
  \label{hex9}
\end{subfigure}
\begin{subfigure}{.23\textwidth}
  \centering
  \includegraphics[width=.999\linewidth]{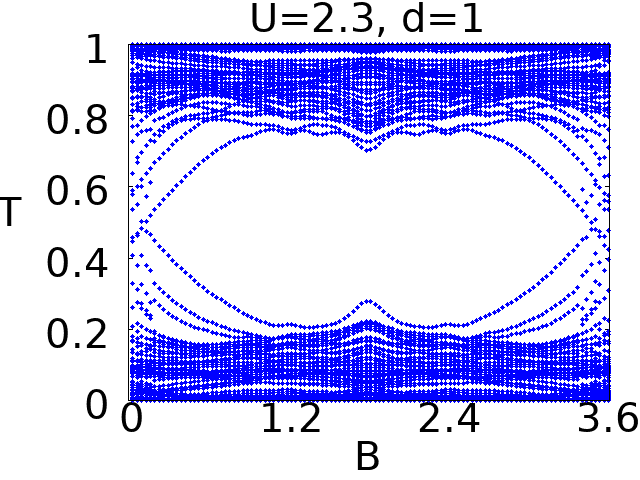}
  \label{hex12}
\end{subfigure}
\caption{The correlation function spectra (T) is plotted against magnetic field(B) for largest subsystem in honeycomb lattice. The notable feature is that for pure case the reflection symmetry exists which is smeared out in the presence of disorder. The interaction induces a gap as before and disorder in the presence of interaction reduces the gap. }
\label{hccorre}
\end{figure}

Now we discuss the entanglement properties for honeycomb lattice. In Fig. \ref{hcentropy} and Fig. \ref{hccorre} we plotted entanglement entropy and correlation spectrum respectively. The Entanglement entropy shows similar characteristics as in square lattice. For pure Hofstadter
 spectrum( in the absence of interaction as well as disorder) the entanglement entropy is equispaced at a given magnetic field. Though the spacing's between two successive subsystem varies with magnetic field. The equal spacing between the entanglement entropy  for various subsystem signifies area law. The entanglement entropy for large subsystem shows a  very weak fluctuation as we increase the magnetic field but it can never be termed as oscillations as found for square lattice.  We find that for honeycomb lattice the departure from area law for larger subsystem does not exist  contrary to  square lattice. In Fig. \ref{hc-area-ipr-edge} upper left and right panel we plotted entanglement entropy versus L for three different magnetic field for uniform and disordered case respectively. The entanglement entropy for higher magnetic fields are shifted vertically by 7.5 and 15 for better representation. Perfect linear behavior of the plot shows agreement to area law. In the middle left and right panel we showed IPR and we see that disorder has induced an increment in IPR by a factor of 10. In the lower panels of Fig. \ref{hc-area-ipr-edge}, we plotted amplitude of the wave functions(from the middle of one of the edge to the center) at half filing and we observe the quasi-delocalized nature of the wave function similar to square lattice.  Though the effect of disorder induces an increment of IPR by 10 times for both square and honeycomb lattice, we observe that the area law is better maintained for honeycomb lattice.  This probably indicates that the wave functions are more homogeneous and uniform in case of honeycomb lattice such that instead of quasi-delocalized modes a better adherence to area law is found in comparison to square lattice.  For large subsystem, the entanglement entropy monotonically increases with intermittent  minima including a  minima  around $B=\pi/\sqrt{3}$ which is half-cycle values for the magnetic field. In the presence of interaction and disorder, some feature of only disorder case returns and also net entanglement entropy increases.  The entanglement spectrum  also shows interesting  pattern similar to square lattice. For pure case i.e in the absence of interaction and disorder, the spectrum do have a reflection symmetry and finer structure as found  in the Hofstadter spectrum.  In the presence of disorder the spectrum becomes more homogeneous without any finer structure. In the presence of only interaction a large gap opens in the spectrum. This gap is minimum at $B=0, 2 \pi/\sqrt{3}$, i.e at the zone boundary. For the square lattice at the zone boundary the interaction induced gap is maximum.   When disorder is turned on in the presence of interaction we find that interaction induced gap is reduced and at the zone boundary the gap almost vanishes. Thus we see the signature of nullifying effect of interaction and disorder in the correlation spectrum as well.

\section{Discussion}
\label{discussion}
Let us recapitulate and summarize our results briefly. As already explained  the main objective of our study is to explore the interplay of disorder
and interaction in the Hofstadter spectra and entanglement as well.  The disorder is implemented according to
four letter Rudin-Shapiro sequence as we intended to see the effect of disorder as close as possible to random disorder without doing configurational average. However, it may remain to  justify such claim of simulating the actual physics obtained through configurational averages over many random disordered realizations. Nevertheless given the fact that such configurational averages needs a much longer time to investigate, the Rudin-Shapiro sequence followed in this article always, on its own, constitute a particular realizations of disorderedness which deserves to be explored. We note that such method of investigations which avoid configuational averages are present in condensed matter system such as Fibonacci series \cite{vipin1}, Aubry-Andre-Harper model\cite{vipin2} .  
To account for the interaction, we performed a self consistent unrestricted Hartree-Fock analysis.
 As expected the butterfly nature of the spectra vanishes once the disorder is turned on in the absence of interaction. When the interaction is present 
 but the disorder is not present, we find that a gap comes to exist in the middle of the Hofstadter spectrum but otherwise the pattern remains intact as found before ~\cite{Cjazka,Minh}.
 Remarkably, in simultaneous presence of disorder and interaction, we find that
 butterfly nature of the Hofstadter spectrum returns to some extent with the wings having lesser width.
Also the interaction induced gap in the middle of the Hofstadter spectra vanishes.  This implies that interaction
 and disorder successfully nullify each other.  It would be interesting to see what happens to such  revival beyond mean field. However this is beyond the scope of present study.  The revival of Hofstadter butterfly intricately depends on the lattice 
coordination number as well as the relative strength of disorder. For example,
 effect of disorder is  enhanced in the honeycomb lattice because of lower
 coordination number. This is manifested in the comparatively weaker revival of butterfly spectra in Honeycomb lattice than in square lattice.  Also the effect of  
disorder is nullified by interaction at  $U=1.9$ for square lattice whether for honeycomb lattice the corresponding value is $U=2.3$. If we increase $U$
 further beyond $U= 2.3$, a gap opens up at Half-filling, therefore this is the farthest we can go for nullification.  \\
\indent
The interplay of interaction and disorder has also been successfully realized in the study of entanglement properties. In our endeavor to study the entanglement properties of the Hofstadter effect, we find that for  pure case, area law of entanglement entropy is preserved for
low and high magnetic field. But for intermediate magnetic field the entanglement entropy depart from its usual area law for square lattice for larger subsystem. For honeycomb lattice there is perfect agreement with area law in all cases. Our effort to understand this differences between square lattice and honeycomb lattice through investigating various system sizes and different disorder realizations point out that there is intricate differences in the resulting wave functions in square and honeycomb lattice. The inverse participation ratio as well as distributions of wave functions show the existence of large number of quasi-delocalized modes which are possibly fundamentally different in square and honeycomb lattice. Possibly for larger subsystems such wave functions are at maximum incommensurability resulting such violations of are law. But for smaller subsystems the effect is minimized. Finally we see that the correlation function eigenvalues do manifest the symmetric properties observed in the Hofstadter butterfly. The effect of interaction does recover
 the area law at intermediate magnetic field implying the localization of the eigenstates due to interactions are in effect.  In the presence of disorder, we similarly find that area law is better recovered for square lattice. In the entanglement spectrum as well we find that the symmetry of the spectra is lost and becomes homogeneous in the presence of disorder, though the spectrum is more denser at higher and lower limits of the value of correlation function spectra. The effect of interaction on the entanglement entropy and spectrum is also manifested profoundly for the intermediate magnetic field. The gap of the Hofstadter spectrum due to interaction are also reflected   in the gap of the correlation (entanglement spectrum) spectrum. These band gaps are minimum where the gap of the Hofstadter spectra are minimum in the presence of interaction \cite{fidkowski-2010}. The entanglement entropy as well as entanglement spectra does show similar behavior in honeycomb lattice.  \\
 \indent
Thus our study of the effect of interaction and disorder on square and honeycomb lattice in view of the modification of Butterfly effect and entanglement shows 
importance of lattice connectivity to determine the ground state properties. Though we have used self consistent mean field approximation, we think that a
more accurate calculation would not qualitatively change the finding of our study. However it would be interesting to derive various aspect of our
 study for example the value of critical interaction strength and disorder at the nullification point in an analytical way. However such study is beyond the scope of the present study and will be carried out in future. We may note that the mean-field  method has been very effective in capturing the transition from charge density wave to spin density wave phase for  one dimensional chemically modulated Hubbard chain at half filling and there was excellent agreement with the real space re-normalization method results~\cite{gupta3}. The transition from paramagnetic to antiferromagnetic phase in two dimensional  $t-t^{\prime}$  Hubbard model at half filling was captured very well by the UHF method and was in close agreement with the quantum Monte Carlo results\cite{lin-1987}.  Whereas in case of correlated Kondo type processes where there is a large effective mass re-normalization near Mott transition, methods like dynamical mean field theory are more capable than UHF in capturing the transition~\cite{potthoff,lederer}. In light of this we think that the the parameter space being implemented in the present study is a realistic one. We think that with the advancement of recent progresses in cold atom systems in optical lattices \cite{sen-sen-2007,joller-2003,peden,osterloh} effect of interaction and disorder on Hofstadter butterfly may be achieved. Further in view of the scope of future avenue it is interesting to note that there are  previous studies~\cite{callan-1992,ouvry} which  investigated the effect of dissipation \cite{callan-1992} as well as non-hermitian effect \cite{ouvry} on Hofstadter effect. As the dissipation parameterized by the friction parameter $\alpha$ become greater than unity one obtains a phase transition from delocalised phase to localized phase. On the other hand the non-hermitian effect as modeled by the biased hopping along $x$-direction shows a nested multi fractal structure. It may be noted that recently a lot of interest is shown in investigating the non-hermitian effect~\cite{ashida-2020,ramy-2017} in various condensed matter system and its relation to  dissipation and interaction. Thus our present study could be further extended in these directions and will be taken as a future endeavor.
\section*{Acknowledgement}
S M thanks SAMKHYA: High Performance Computing Facility provided by the Institute of Physics, Bhubaneswar. S M also acknowledges S P Mandal without whose hospitality it would be impossible to complete the project in time.
\section*{Appendix}
In this appendix we briefly explain and clarify various associated definitions and formulas that have been used in the text. We begin by a brief introduction of entanglement entropy. \\
\indent
{\bf Entanglement entropy for non-interacting tight binding Hamiltonian} It is known that entanglement is understood mainly between two
quantum states where two states could be qubits or representative of two different region in space and time. Here we are interested in the entanglement between
two regions of interest say A and B. This two regions comprise the full system which in our case is a finite lattice such as square or honeycomb. The first step toward such estimation of entanglement entropy is calculation of reduced density matrix of A or B. For a tight binding model having only hopping interaction such as $H=\sum _{i,j}t_{ij}c^{\dagger}_i c_j $, the reduced density matrix of a given region say A can be calculated exactly by calculating the correlation function matrix of that region. By correlation  function matrix we mean that if the region A contains $N_A$ sites the correlation function matrix is   $N_A \times N_A$ matrix i.e $\mathcal{C}= <c^{\dagger}_{i^{\prime}} c_{j^{\prime}}>$ where $i^{\prime}$ and $j^{\prime}$ denote any two generic sites inside the region $A$. For a free fermionic Hamiltonian as $H$, one can exactly calculate the eigenvalues of $\mathcal{C}$ as described in \onlinecite{ent}. There is one to one correspondence between the eigenvalues of the correlation function matrix (say $\lambda_m, m=1, N_A$) and the eigenvalues of the reduced density matrix of that region.  By Wick's theorem and following reference \onlinecite{ent}, one can write the reduced density matrix of the region A in the following form,
\begin{eqnarray}
\label{rhored}
\rho_A&=& \mathcal{K} e^{-\sum_k \epsilon_k n_k} .
\end{eqnarray}
In the above $n_k$'s are fermionic  number operators corresponding to some fictitious eigenvalues $\epsilon_k$'s which are related to the eigenvalues of the correlation matrix eigenvalues $\lambda_k$ by,
\begin{eqnarray}
\epsilon_k=log \frac{1-\lambda_k}{\lambda_k} .
\end{eqnarray}

It is easy to find the constant in Eq. \ref{rhored} from the condition $Tr(\rho_A)=1$ and one obtains $\mathcal{K}=\prod^{N}_{k=1} (1-\lambda_k)$. Given the $N_A$ eigenvalues $\epsilon_k$, one can then easily obtain the $2^{N_A}$ eigenvalues of the reduced density matrix from Eq. \ref{rhored}.  These eigenvalues are expressed as,
\begin{eqnarray}
 \Lambda=   \prod   (1-\lambda_k)^{1-n_k} \lambda^{n_k}_{k}, ~~~ n_k=0,1
\end{eqnarray}
There are $2^{N_A}$ choices of configurations of $n_k$ which yield the total $2^{N_A}$ eigenvalues.
The von-Neuman entanglement entropy $\mathcal{S}_{en}$ is defined as,

\begin{eqnarray}
\mathcal{S}_{en}=- Tr \left( \rho log \rho  \right)
\end{eqnarray}

Using the definition of $\rho$ as given in Eq. \ref{rhored}, one arrives at the final formula for entanglement entropy as,

\begin{eqnarray}
\mathcal{S}_{en}&&= \sum_{k} -\left(\lambda_k log \lambda_k+ (1-\lambda_k) log (1-\lambda_k)   \right)
\end{eqnarray}

Thus we observe that entanglement entropy is completely determined by the eigenvalues of the correlation function matrix $\mathcal{C}$. Apart from the entanglement entropy entanglement spectra i.e distribution of $\epsilon_k$ is also found to be important. Instead of $\epsilon_k$, one can equally look at correlation  matrix spectra i,e the distribution of $\lambda_k$. In this study we have numerically studied entanglement entropy as well as correlation function spectra. \\
\indent

\begin{figure}[!htb]
\begin{subfigure}{.23\textwidth}
  \centering
  \includegraphics[width=.99\linewidth]{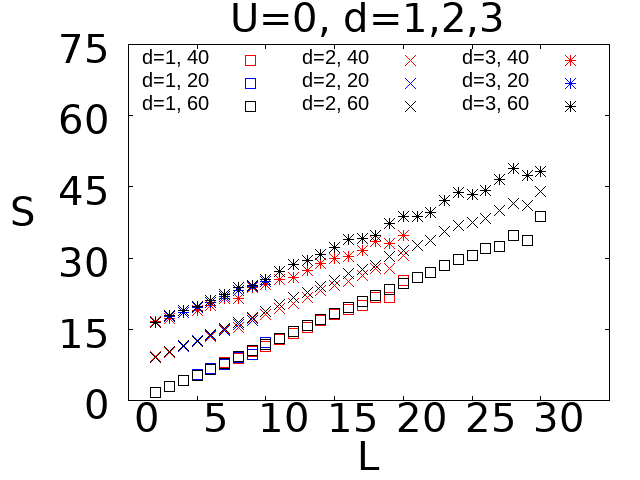}
  \label{dsq1}
\end{subfigure}
\begin{subfigure}{.23\textwidth}
   \centering
 \includegraphics[width=.99\linewidth]{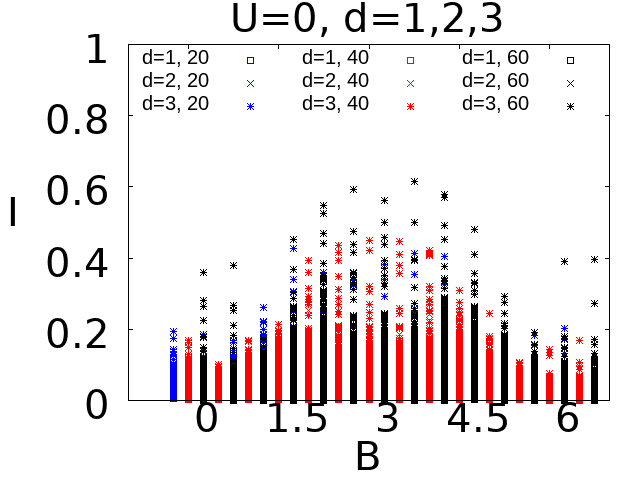}
\label{dsq-ipr}
\end{subfigure}
\begin{subfigure}{.23\textwidth}
  \centering
  \includegraphics[width=.99\linewidth]{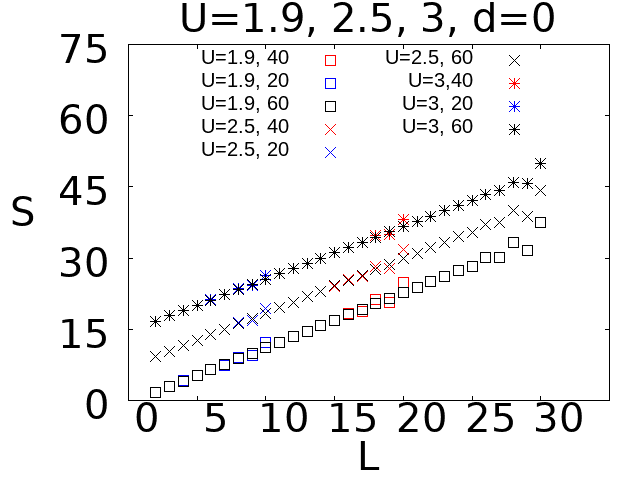}
  \label{dsq2}
\end{subfigure}
\begin{subfigure}{.23\textwidth}
   \centering
 \includegraphics[width=.99\linewidth]{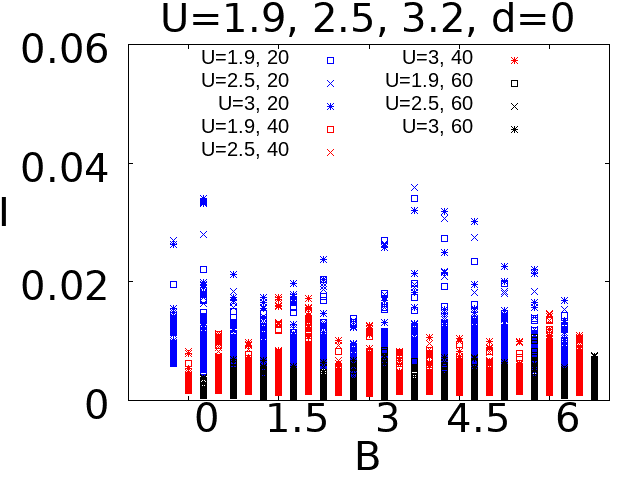}
\label{usq-ipr}
\end{subfigure}
\caption{In the upper left panel entanglement entropy has been plotted  against subsystem sizes for various system size and disordered realization. For disorder $d=2, 3$ the entanglement entropy is vertically shifted by 5 and 10 units for visual clarity. In the lower left panel we plotted the entanglement entropy for three different realizations of interaction and similar convention has been followed as before. In the right upper panel and lower panel inverse participation ratio for various realization of disorder and for system size with $L= 20, 40 $ and 60.  The IPR for system size 20 and 60 is shifted  by $\pm 0.5$ unit vertically for better visualization. In the lower right panel we plotted IPR for various realization of interaction for system sizes 20, 40 and 60. We followed same convention as in the upper right panel. }
\label{sqarealaw}
\end{figure}

{\bf Notes on finite size effect of entanglement entropy:} Now we discuss the effect of subsystem size and strength of disorder as well as of  interaction.  In the upper left panel of Fig. \ref{sqarealaw}, we have plotted entanglement entropy for three different realizations of disorders.  Also for each realizations, we took three different system sizes such as $20\times 20$, $40\times 40$ and $60 \times 60$. The subsystem for each of these three systems begin with $2 \times 2$ and ends at $L/2  \times L/2$ where $L=20,40,60$. As can be seen that for system sizes with $L=20$ the largest subsystem does not show much departure from the linear behavior. However  for system sizes 40 and 60, the larger subsystems always show some departure from linear behavior i.e from area law. If there is a genuine boundary effect which is causing these departure one would expect that such effect to disappear with larger system and subsystem sizes.  This must come from some quasi-delocalized or fully delocalized modes that still exist in the presence of disorder. In the right lower panel of Fig. \ref{sqarealaw}, we have plotted entanglement entropy for three different realizations of interactions and again we observe that there is departure from linear behaviour for larger subsystems independent of full system sizes. This also excludes the existence of any boundary effect and instead points out to the complexity of bulk modes at half filling. In the right panels we plotted inverse participation ratios (IPR) for disorder (upper panel) and interaction (lower panel). It shows that disorder is more effective into inducing localization as the IPR for interaction as lower by 10 times compared to what is obtained in the presence of interaction. However the departure from area law for the larger subsystems as discussed before indicates that eigenstates with low IPR values (which are possibly well delocalized) are causing this.
\\
\indent

 {\bf Definition of Inverse Participation Ratio:} Consider that a given single particle normalized wavefunction $| \Psi \rangle$ can be expanded in the following way,
\begin{eqnarray}
|\Psi \rangle && = \sum_i a_i c^{\dagger}_i |0\rangle
\label{psiipr}
\end{eqnarray}
In the above $c^{\dagger}_i$ is the creation operator at site `$i$' and $a_i$ denotes the probability amplitude. The inverse participation ratio~\cite{maestro,romera} that is used to describe the degree of localization is defined as,

\begin{eqnarray}
I&&= \sum_i |a_i|^4
\end{eqnarray}

If the wave function is equally distributed over all the sites, one would have $a_i \approx \frac{1}{\sqrt{N}}$ which yields, $I \approx \frac{1}{N} \rightarrow 0$ for thermodynamic system. On the other hand if the system is sharply localized at a given site say '$j$', the one would have $a_j=1,~~a_{i \ne j}=0$ and $I=1$.
Thus a value of $I$ close to zero describes a delocalized states and if the value of $I$ is close to 1 it describes a localized states. \\

\indent
{\bf Notes on amplitude of wave function:} We refer the definition of $|\Psi \rangle$ as given in Eq. \ref{psiipr}. Now as shown in Fig \ref{fig:ent-subsystem}, the green line describes a path from the boundary to the the center of the finite system that we consider. "S" and "E" denote the starting and end point of this
path. The probability amplitude of $a_i$ ($|a_i|^2$) is plotted along this path in Fig. \ref{sq-area-ipr-edge}, \ref{hc-area-ipr-edge}. The sites along this path start from S (having numbered as 1) and increase upto $N/2$ ending at "E". Our system size is $N \times N$. \\

\begin{figure}[!htb]
\begin{subfigure}{.23\textwidth}
  \centering
  \includegraphics[width=.99\linewidth]{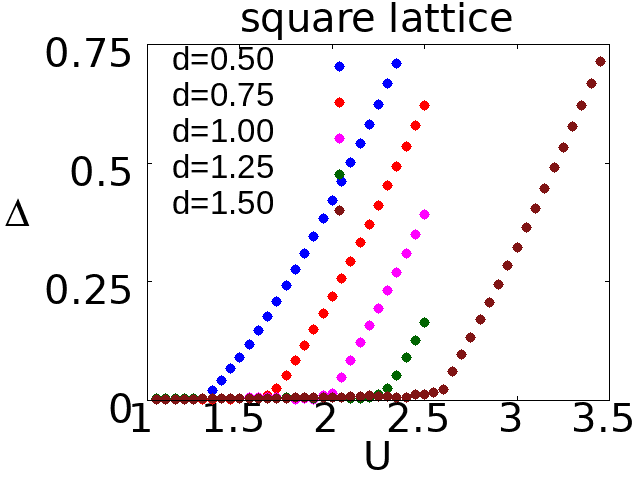}
  \label{sq-cr-u}
\end{subfigure}
\begin{subfigure}{.23\textwidth}
   \centering
 \includegraphics[width=.99\linewidth]{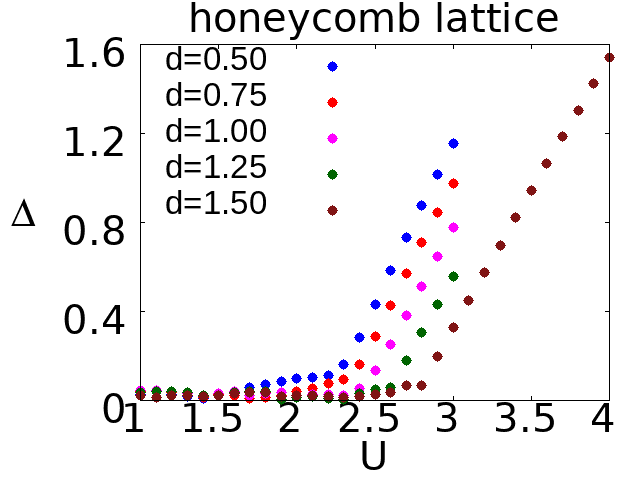}
\label{hc-cr-u}
\end{subfigure}
\caption{In the left panel gap at half filling  for square lattice has been plotted for many disorder strength as denoted. The horizontal axis denotes interaction strength. The right panel shows the gap at half filling for honeycomb lattice. }
\label{ma-ds-ucr}
\end{figure}
\indent
{\bf Discussions on critical interaction strength for various disorders:} A pertinent and very relevant question is what happens if one varies the strength of Rudin-Shapiro
disorder sequence. In the present study the four letter Rudin-Shapiro (A,B,C,D) has been taken as (0,1,0.5,1.5). In the manuscript this disorder is denoted as $d=1$. For this strength critical values of interaction for square and honeycomb lattice are obtained as 1.9 and 2.3 respectively.  Here we briefly discuss what happens if we scale the strength of Rudin-Shapiro disorder sequence  and how the gap at half filling behaves as a function of $U$. The result has been plotted in Fig. \ref{ma-ds-ucr}. As can be seen that  the critical values of $U$ is directly proportional to the applied disordered strength. This further strengthen the main finding of this study is that disorder and interaction really try to nullify each other consistently for a wide range of parameters. From Fig. \ref{ma-ds-ucr} we also note that as disorder strength varies from $0.5$ to $1.50$ the critical value of $U$ is spread from $1.2$ to $2.5$ for square lattice where as for honeycomb lattice it is spread within a narrow region of $2.1$ to $2.9$. Thus we observe that effect of lattice coordinates play a major role in determining the critical values of interaction. However we must note that for various values of $d$ and the corresponding  $U_{cr}$, the disappearance of Hofstadter spectra as well as re-appearance of it would qualitatively change from whatever presented in this study.

\end{document}